\documentclass[twocolumn,showpacs,aps,prd]{revtex4-1}

\usepackage{graphicx}
\usepackage{dcolumn}
\usepackage{amsmath}
\usepackage{epsfig}
\usepackage{subfigure}
\usepackage{epstopdf}
\usepackage{multirow}
\usepackage{mathrsfs}
\usepackage[dvipdfmx, bookmarksnumbered, pdfstartview=FitH,colorlinks,urlcolor=blue, citecolor=blue,linkcolor=blue,] {hyperref}

\usepackage{lineno}
\usepackage{ulem}
\usepackage{enumerate}
\usepackage{float}

\begin{document}


\title{\boldmath Observation of the decays $\chi_{cJ} \to \phi \phi \eta$}

\author{
\begin{small}
\begin{center}
M.~Ablikim$^{1}$, M.~N.~Achasov$^{10,d}$, P.~Adlarson$^{63}$, S. ~Ahmed$^{15}$, M.~Albrecht$^{4}$, M.~Alekseev$^{62A,62C}$, A.~Amoroso$^{62A,62C}$, F.~F.~An$^{1}$, Q.~An$^{59,47}$, Y.~Bai$^{46}$
% ...
\\
\vspace{0.2cm}
(BESIII Collaboration)\\
\vspace{0.2cm} {\it
$^{1}$ Institute of High Energy Physics, Beijing 100049, People's Republic of China\\
% ...
$^{4}$ Bochum Ruhr-University, D-44780 Bochum, Germany\\
% ...
$^{10}$ G.I. Budker Institute of Nuclear Physics SB RAS (BINP), Novosibirsk 630090, Russia\\
% ...
$^{15}$ Helmholtz Institute Mainz, Johann-Joachim-Becher-Weg 45, D-55099 Mainz, Germany\\
% ...
$^{46}$ Southeast University, Nanjing 211100, People's Republic of China\\
$^{47}$ State Key Laboratory of Particle Detection and Electronics, Beijing 100049, Hefei 230026, People's Republic of China\\
% ...
$^{59}$ University of Science and Technology of China, Hefei 230026, People's Republic of China\\
% ...
$^{62}$ (A)University of Turin, I-10125, Turin, Italy; (B)University of Eastern Piedmont, I-15121, Alessandria, Italy; (C)INFN, I-10125, Turin, Italy\\
$^{63}$ Uppsala University, Box 516, SE-75120 Uppsala, Sweden\\
% ...
$^{d}$ Also at the Novosibirsk State University, Novosibirsk, 630090, Russia\\
% ...
}\end{center}

\vspace{0.4cm}
\end{small}}

\begin{abstract}
Using a data sample of $(448.1\pm2.9)\times10^{6}$ $\psi(3686)$ decays collected by the BESIII detector at the Beijing Electron Positron Collider (BEPCII), we observe the decays $\chi_{cJ}\to \phi\phi\eta~(J=0,~1,~2)$, where the $\chi_{cJ}$ are produced via the radiative processes $\psi(3686)\to\gamma\chi_{cJ}$. The branching fractions are measured to be $\mathcal B(\chi_{c0}\to\phi\phi\eta)=(8.41\pm0.74\pm0.62)\times10^{-4}$,
$\mathcal B(\chi_{c1}\to\phi\phi\eta)=(2.96\pm0.43\pm0.22)\times 10^{-4}$, and $\mathcal B(\chi_{c2} \to \phi\phi\eta)=(5.33\pm0.52\pm0.39) \times 10^{-4}$, where the first uncertainties are statistical and the second are systematic. We also search for intermediate states in the $\phi\phi$ or $\eta\phi$ combinations, but no significant structure is seen due to the limited statistics.

\end{abstract}


\maketitle

\section{\boldmath INTRODUCTION}

 Studies of the properties of $c\bar{c}$ states play an important role in understanding the interplay between perturbative and non-perturbative effects in quantum chromodynamics (QCD). Besides $J/\psi$ and $\psi(3686)$ decays~\cite{pdg}, the decays of the $\chi_{cJ}$ ($J=0,~1,~2$)~\cite{chicj_first_1,chicj_first_2} are also valuable to probe a wide variety of QCD phenomena.

To date, only a few measurements have been performed for decays of the form $\chi_{cJ}\to VVP$, where $V$ and $P$ denote vector and pseudoscalar mesons, respectively~\cite{pdg}, and no measurement of the branching fraction for $\chi_{cJ}\to\phi \phi \eta$ has previously been reported. The interest in these final states arises from the search for glueballs in the $\phi\phi$ invariant mass ($M_{\phi\phi}$) spectrum. A previous partial wave analysis of the decay $J/\psi \to \gamma \phi \phi$ decay by the BESIII Collaboration~\cite{aixiaocong} confirmed the existence of the $\eta(2225)$ and observed the three tensor states $f_2(2010),~f_2(2300)$ and $f_2(2340)$, which were first observed in the process $\pi^- p \to \phi\phi n$~\cite{phiphi_ppi}. Different experiments also searched for glueballs~\cite{glueball_B} decaying to $\phi\phi$ in $B$ decays, but none have so far been observed~\cite{pdg}. Although there are no theoretical expectations, the decays $\chi_{cJ}\to\phi\phi\eta$ may contain contributions from intermediate states decaying to $\phi\phi$ and $\eta\phi$, and observations of the same resonances as those in $J/\psi$ decays would provide supplementary and conclusive information regarding their existence.

Due to abundant $\chi_{cJ}$ production in $\psi(3686)$ radiative decays~\cite{pdg}, the BESIII experiment provides an ideal place to search for new $\chi_{cJ}$ decays based on the world's largest $e^+ e^-$ annihilation data sample of $(448.1\pm2.9)\times10^{6}$ $\psi(3686)$ events~\cite{psipnumberT}. In this paper, we report the first measurements of the branching fractions of $\chi_{cJ}$ decays to $\phi \phi \eta$. The $\phi$ meson can be reconstructed with $\phi \to K^+ K^-$, $\phi \to \pi^+ \pi^- \pi^0$ and $\phi\to K_S^0 K_L^0$ decays, and the $\eta$ meson with $\eta \to \gamma \gamma$ and $\eta\to \pi^+ \pi^- \pi^0$ decays. Compared to the $\phi\to K^+ K^-$ and $\eta \to \gamma \gamma$ modes, other decay modes suffer from higher backgrounds and lower detection efficiencies. So in this analysis, the two $\phi$ mesons and the $\eta$ meson are reconstructed with $\phi \to K^+ K^-$ and $\eta \to \gamma \gamma$ processes.

\section{\boldmath{Detector and Monte Carlo Simulations}}
The BESIII detector is a magnetic
spectrometer~\cite{Ablikim:2009aa} located at the Beijing Electron
Positron Collider (BEPCII)~\cite{Yu:IPAC2016-TUYA01}. The
cylindrical core of the BESIII detector consists of a helium-based
 multilayer drift chamber (MDC), a plastic scintillator time-of-flight
system (TOF), and a CsI(Tl) electromagnetic calorimeter (EMC),
which are all enclosed in a superconducting solenoidal magnet
providing a 1.0~T magnetic field. The solenoid is supported by an
octagonal flux-return yoke with resistive plate counter muon
identifier modules interleaved with steel. The acceptance for charged particles and photons is 93\% over $4\pi$ solid angle. The
charged-particle momentum resolution at $1~{\rm GeV}/c$ is
$0.5\%$, and the $dE/dx$ resolution is $6\%$ for the electrons
from Bhabha scattering. The EMC measures photon energies with a
resolution of $2.5\%$ ($5\%$) at $1$~GeV in the barrel (end cap)
region. The time resolution of the TOF barrel part is 68~ps, while
that of the end cap part is 110~ps.

Large samples of simulated events are produced with a {\sc
geant4}-based~\cite{geant4} Monte Carlo (MC) package that
includes the geometric description of the BESIII detector and the
detector response. These samples are used to determine the detection efficiency
and to estimate the backgrounds. The simulation includes the beam
energy spread and initial state radiation (ISR) in the $e^+e^-$
annihilation modeled with the generator {\sc
kkmc}~\cite{ref:kkmc}. The `inclusive' MC sample consists of the production of the
$\psi(3686)$ resonance, the ISR production of the $J/\psi$, and
the continuum processes incorporated in {\sc
kkmc}~\cite{ref:kkmc}. The known decay modes are modeled with {\sc
evtgen}~\cite{ref:evtgen} using branching fractions taken from the
Particle Data Group~\cite{pdg}, and the remaining unknown decays
of the charmonium states are modeled with {\sc
lundcharm}~\cite{ref:lundcharm}. The final state radiation (FSR)
from charged final state particles is simulated with the {\sc
photos} package~\cite{photos}.

 For the signal MC samples, the $\psi(3686) \to \gamma \chi_{cJ}$ decays are generated with the electric dipole (E1) transition~\cite{E1_1, E1_2} assumption, where the angular distribution is $1+\lambda \cos^2 \vartheta$~\cite{E1_form1, E1_form2}. Here, $\vartheta$ is the polar angle of the radiative photon in the rest frame of the $\psi(3686)$ meson, and $\lambda$ is $1$, $-1/3$, $1/13$ for $J=0$, $1$, $2$, respectively. The processes $\chi_{cJ}\to \phi \phi \eta$ and $\eta\to\gamma \gamma$ are generated uniformly in phase space, and the angular distribution of the $\phi\to K^+ K^-$ decay is modeled as a vector particle decaying to two pseudoscalars.

\section{\boldmath{Event Selection}}

The cascade decay of interest is $\psi(3686) \to \gamma \chi_{cJ},~\chi_{cJ} \to \phi \phi \eta$, with $\phi \to K^+ K^-$ and $\eta \to \gamma \gamma$. Candidate events are required to have four charged tracks with zero net charge and at least three photons. Charged tracks in an event are required to have a polar angle $\theta$ with respect to the beam direction within the MDC acceptance $|\cos \theta|<0.93$, and a distance of closest approach to the interaction point within 10 cm along the beam direction and 1 cm in the plane transverse to the beam direction. The TOF and $dE/dx$ information are combined to evaluate particle identification (PID)
 confidence levels for the $\pi$ and $K$ hypotheses, and the particle type with the highest confidence level is assigned to each track. All charged tracks must be identified as kaons. Electromagnetic showers are reconstructed from clusters of energy deposited in the EMC. The energy
deposited in nearby TOF counters is included to improve the reconstruction efficiency and energy resolution. Photon candidates must have a minimum energy of 25~MeV in the barrel region ($|\cos\theta|<0.80$) or 50~MeV in the end cap region ($0.86<|\cos\theta|<0.92$).
To exclude showers from charged particles, a photon must be separated by at least $10^\circ$ from the nearest charged track. The measured EMC time is required to be within 0 and 700 ns of the event start time to suppress electronic noise and energy deposits unrelated to the event of interest.

 A four-constraint (4C)
kinematic fit imposing overall energy-momentum conservation is performed with the $\gamma\gamma\gamma K^+ K^- K^+ K^-$ hypothesis, and the events with $\chi^2_{\mathrm{4C}}<40$ are retained. The requirement is based on the optimization of the figure of merit (FOM), FOM $\equiv$ $N_{\rm{sig}}/\sqrt{N_{\rm tot}}$ , where $N_{\rm{sig}}$ and $N_{\rm tot}$ are the number of signal events and total number of events estimated from the signal MC sample and data, respectively. For events with more
than three photon candidates, the combination with the smallest $\chi^2_{\mathrm{4C}}$ is retained. Further selection criteria are based on the four-momenta updated by the kinematic fit.

\begin{figure}[tb]
\centering
   \includegraphics[width=0.41\textwidth]{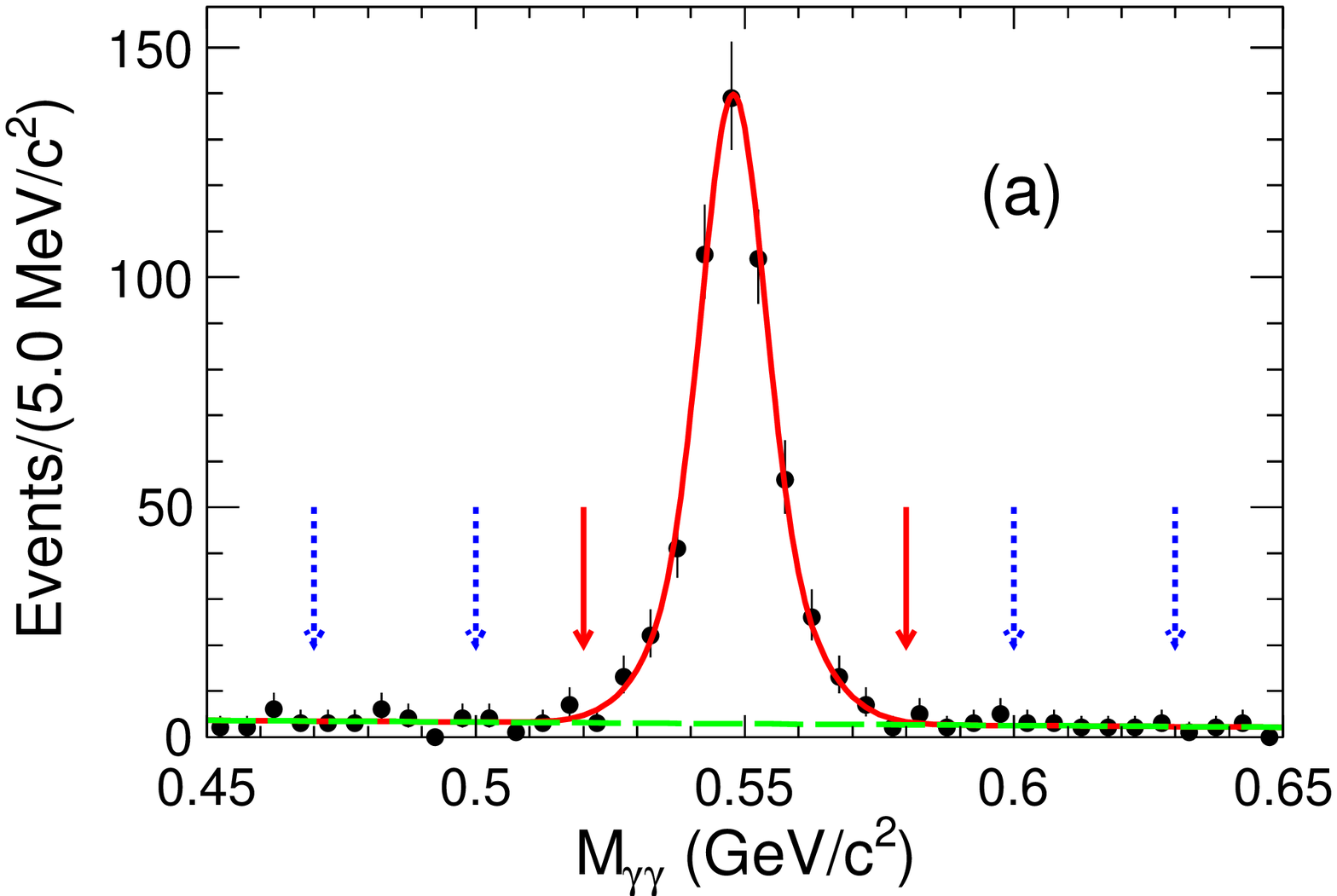}
    \includegraphics[width=0.41\textwidth]{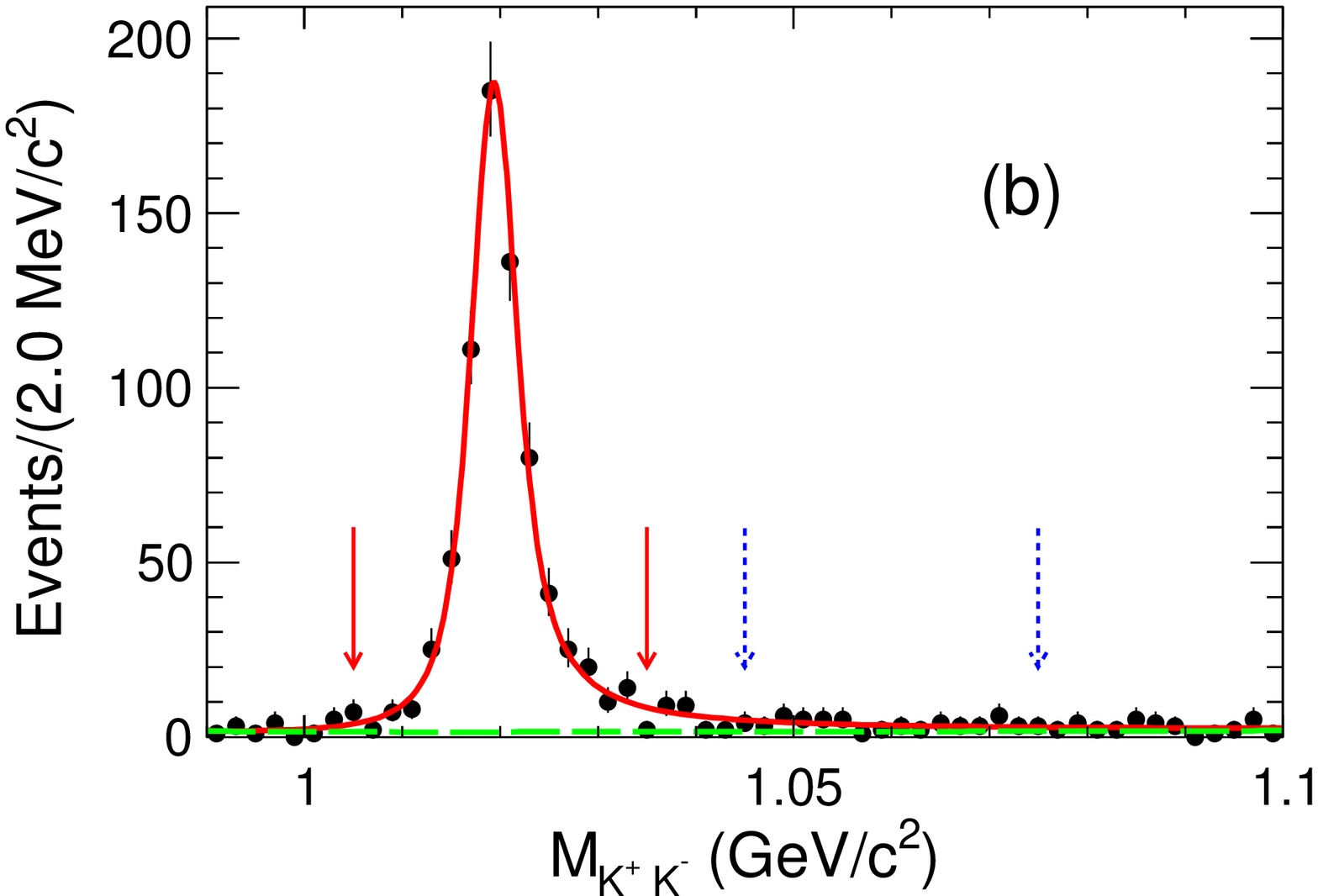}
  \caption{Fits to (a) $M_{\gamma\gamma}$ and (b) $M_{K^+ K^-}$, where one of the two combinations is randomly selected. The dots with error bars are from data, the red lines are the best fit results, and the long dashed green lines are the background shapes. The red arrows show the signal region, and the dashed blue arrows show the sideband regions.}
\label{fit_meta_mphi_mass}
\end{figure}

\begin{figure}[tb]
\centering
   \includegraphics[width=0.45\textwidth]{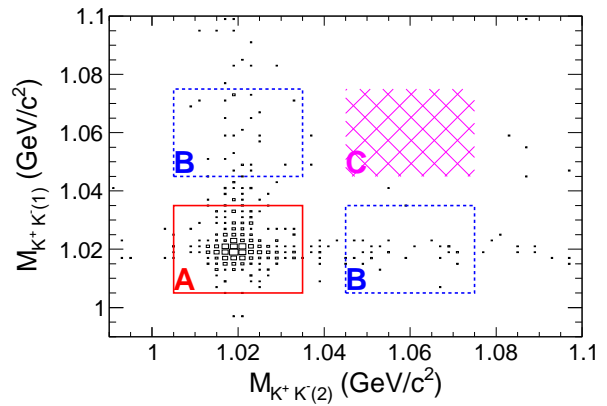}
  \caption{Distribution of $M_{K^+ K^-(1)}$ versus $M_{K^+ K^- (2)}$. The solid red rectangle (area ``A") and dashed blue rectangles (areas ``B") denote the 2-D $\phi$ signal region and 2-D $\phi$ sideband regions, respectively, and the hatched pink rectangle (area ``C") means both of $M_{K^+ K^-(1)}$ and $M_{K^+ K^- (2)}$ lie in the $\phi$ sideband region.}
\label{mphi_2d}
\end{figure}

After the above requirements, the $\eta$ candidate is reconstructed in its decay to $\gamma\gamma$ using the $\gamma \gamma$ pair with invariant mass $M_{\gamma \gamma}$ closest to the nominal $\eta$ mass~\cite{pdg}. The $\eta$ signal
region is defined as $0.52~\leq~M_{\gamma \gamma}~\leq~0.58$~GeV/$c^2$, with half-width approximately three times larger than the detector resolution ($\sigma_{\eta}=10$~MeV/$c^2$). Figure~\ref{fit_meta_mphi_mass}(a) shows a fit to the $M_{\gamma \gamma}$ distribution. In the fit, the signal shape is modeled by the MC-simulated lineshape convolved with a Gaussian function with free width and the background is described by a linear function. The two signal $\phi$ candidates are chosen from the combination with the minimum value of $\Delta M^2 = (M_{K^+_i K^-_j} - m_{\phi})^2 + (M_{K^+_{1-i} K^-_{1-j}} - m_{\phi})^2$, where $M_{K^+K^-}$ is the invariant mass of $K^+ K^-$, $m_{\phi}$ is the nominal $\phi$ mass~\cite{pdg}, and $i,j$ can be 0 or 1. MC studies show that the miscombination rates for both $\eta$ and $\phi$ candidates are no more than 0.1\%. The $\phi$ signal region is defined as $1.005~\leq~M_{K^+ K^-}~\leq~1.035$~GeV/$c^2$, with half-width about three times the sum of the detector resolution ($\sigma_{\phi}=1$~MeV/$c^2$) and intrinsic width~\cite{pdg}. Figure~\ref{fit_meta_mphi_mass}(b) shows the fit to the $M_{K^+ K^-}$ distribution obtained when one of the two $\phi$ candidates is randomly selected. In the fit, the signal shape is modeled as a $P$-wave Breit-Wigner convolved with a Gaussian function, and the background shape is represented by the function $b(M_{K^+ K^-})=(M_{K^+ K^-}-m_t)^c e^{-dM_{K^+ K^-}}$, where $m_t$ is the $K^+ K^-$ mass threshold, and $c$ and $d$ are free parameters. The two-dimensional (2-D) $\phi$ signal region is shown as the area ``A" in Fig.~\ref{mphi_2d}, where $M_{K^+ K^-(1)}$ and $M_{K^+ K^-(2)}$ denote the invariant masses of the two $\phi$ candidates.

The mass recoiling against the $\eta$ is required to be less than 3.05~GeV/$c^{2}$ to suppress background from the decay $\psi(3686)\to \eta J/\psi,~J/\psi\to\gamma \phi \phi$. All combinations of $M_{\gamma \gamma}$ are required to be outside the range [0.115, 0.150]~GeV/$c^{2}$ to suppress background events with $\pi^0$ decays, and the invariant mass of $\gamma\eta$ must be outside the range [1.00, 1.04]~GeV/$c^2$ to suppress background from the decay $\psi(3686)\to\phi\phi\phi$, where one $\phi$ decays to $\gamma \eta$.

\begin{figure}[tb]
\centering
\includegraphics[width=0.45\textwidth]{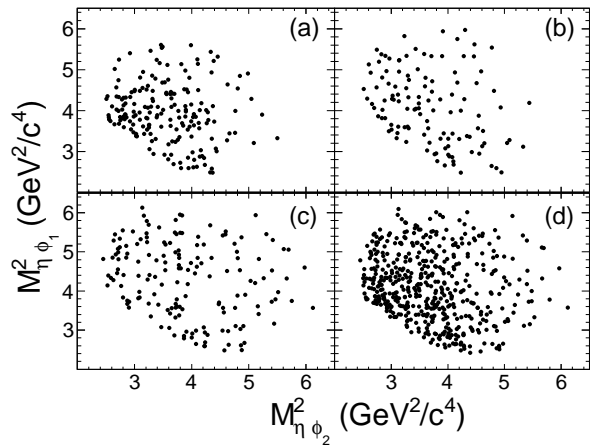}
  \caption{Distributions of $M^2_{\eta\phi_1}$ versus $M^2_{\eta\phi_2}$ for $\chi_{cJ} \to \phi \phi \eta$ decays within the signal regions of the (a) $\chi_{c0}$, (b) $\chi_{c1}$, and (c) $\chi_{c2}$, as well as (d) the overall region [3.3, 3.6]~GeV/$c^{2}$.}
\label{dalitz}
\end{figure}

A total of 495 candidate events survive the event selection criteria. The distributions of $M^{2}_{\eta\phi_{1}}$ versus $M^{2}_{\eta\phi_{2}}$ from the three $\chi_{cJ}$ states are depicted in Fig.~\ref{dalitz}, where the signal regions of $\chi_{c0}$, $\chi_{c1}$, and $\chi_{c2}$ are defined as [3.38, 3.45], [3.48, 3.54], and [3.54, 3.60]~GeV/$c^{2}$ for the invariant mass $M_{\phi\phi\eta}$, respectively.

%


\section{\boldmath{BACKGROUND and Signal yields}}
\label{sectionbkg}

 According to a study of the inclusive MC sample, consisting of $5.06\times 10^8$ $\psi(3686)$ events, the background sources can be categorized into two classes. The class I background is from the decays with no $\eta$ signal formed in the $M_{\gamma \gamma}$ distribution, which is estimated by the events in the $\eta$ mass sideband regions of $M_{\gamma \gamma}\in~$[0.47, 0.50]~$\cup$~[0.60, 0.63]~GeV/$c^{2}$. The class II background arises from the decays with only one $\phi$ signal, which is described with the events in the 2-D $\phi$ sideband regions (the areas ``B" of Fig.~\ref{mphi_2d}), where one $M_{K^+ K^-}$ lies in the $\phi$ signal region and the other is in the $\phi$ sideband region of $M_{K^+ K^-}\in$ [1.045, 1.075]~GeV/${c^2}$. Since there are no events observed in the area ``C" of Fig.~\ref{mphi_2d}, in which both $M_{K^+ K^-}$ lie in the $\phi$ sideband region, we ignore this contribution.

The quantum electrodynamics process under the $\psi(3686)$ peak is studied based on the off-resonance sample of 48.8~pb$^{-1}$ taken at the center-of-mass energy of 3.65~GeV~\cite{paper_hjli}. With the same event selection criteria, no
  events survive, so this contribution is also negligible.

\begin{figure}[tb]
\centering
   \includegraphics[width=0.42\textwidth]{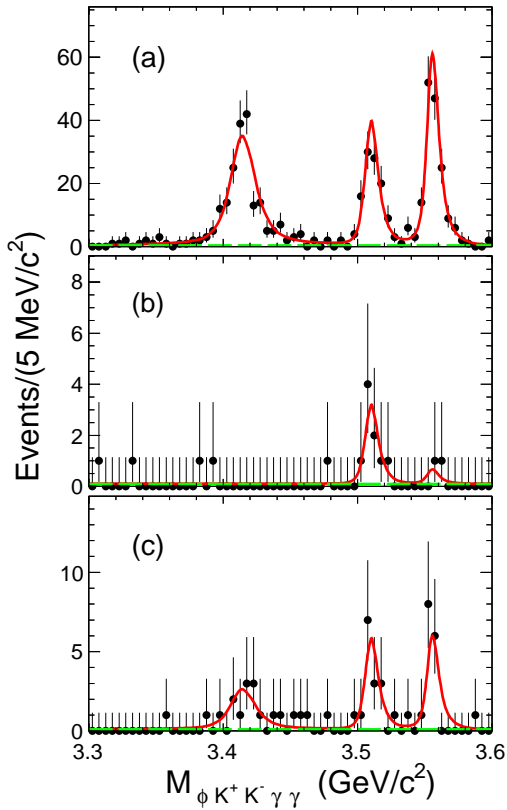}
  \caption{Fits to the $M_{\phi K^+ K^- \gamma \gamma}$ distributions for (a) the ``R1" region, (b) the ``R2" region, and (c) the ``R3" region. The dots with error bars are from data, the solid red lines are the best fit results, and the long dashed green lines are the fitted backgrounds.}
\label{fit_chicj}
\end{figure}

    The signal yields are obtained from unbinned maximum likelihood fits to the $M_{\phi K^+ K^- \gamma \gamma}$ spectra, where at least one of the $\phi$ candidates has the invariant mass within the signal window. The fits are performed in the following three regions, ``R1", ``R2", and ``R3", which correspond to the area ``A" with the $\eta$ in the signal region, the area ``A" with the $\eta$ in the sideband regions, and the areas ``B" with the $\eta$ in the signal region, respectively. In the fits, the signal shape is extracted from signal MC simulations, and the background shape is modeled as a constant. Figure~\ref{fit_chicj} shows the fit results. The contribution of the areas ``B" with the $\eta$ in the sideband region is negligible, since there are only 2 events. The signal yields for $\chi_{cJ} \to \phi \phi \eta$ decays are estimated by
\begin{equation}
  \label{Amp}
  N_{\rm obs}^{\rm sig} = N_{\rm obs}^{\rm R1} -  f_{\rm R2}\cdot N_{\rm obs}^{\rm R2} - f_{\rm R3}\cdot N_{\rm obs}^{\rm R3},
\end{equation}
where $N_{\rm obs}^{r}$ is the number of observed events for the corresponding $r$ region ($r=$ R1, R2, or R3), and both the normalization factors $f_{\rm R2}$ and $f_{\rm R3}$ are 1.0 evaluated from the ratios of the background yields in the $\eta$ and 2-D $\phi$ signal and sideband regions, respectively. The number of events obtained by the fits to $M_{\phi K^+ K^- \gamma \gamma}$ in different regions for $\chi_{cJ}\to \phi \phi \eta$ decays are summarized in Table~\ref{sig_yield_chicj}, along with their statistical significances.

\begin{table}[tb]
\begin{center}
 \caption{The numbers of observed events for different regions in $\chi_{cJ}\to \phi \phi \eta$ decays, as well as their statistical significances (Sig.). The errors are statistical only.}
  \label{sig_yield_chicj}
  \setlength{\extrarowheight}{1.0ex}
  \renewcommand{\arraystretch}{1.0}
  \vspace{0.2cm}
  \begin{tabular}{p{1.0cm}m{1.8cm}<{\centering}m{1.6cm}<{\centering}m{1.6cm}<{\centering}m{0.7cm}<{\centering}}
  \hline\hline
Mode & $N_{\rm obs}^{\rm R1}$&$N_{\rm obs}^{\rm R2}$ &$N_{\rm obs}^{\rm R3}$ & Sig.  \\\hline
$\chi_{c0}$ & $201.2\pm 15.6$ & $0.0 \pm 0.9$ & $14.6\pm 4.7$ & $18\sigma$\\
$\chi_{c1}$ & $108.0\pm11.0 $ & $8.6 \pm 3.1$  & $15.8\pm 4.2$& $10\sigma$ \\
$\chi_{c2}$ & $160.7\pm13.2$  & $1.5 \pm 1.5$ & $15.6 \pm 4.2$& $17\sigma$ \\
  \hline\hline
  \end{tabular}
  \vspace{-0.2cm}
  \end{center}
\end{table}

\section{\boldmath{Branching fractions}}

The branching fractions of $\chi_{cJ}\to \phi \phi \eta$ decays are determined by
\begin{equation}
  \label{br_cal_result}
  \mathcal{B}_{\chi_{cJ}\to \phi \phi \eta} = \frac{N^{\rm sig}_{\rm obs}}{N_{\psi(3686)} \cdot\mathcal{B}_{0}\cdot \epsilon},
\end{equation}
where $N_{\psi(3686)}$ is the total number of $\psi(3686)$ events, $\mathcal{B}_{0} = \mathcal{B}_{\psi(3686)\to\gamma\chi_{cJ}}\mathcal{B}^2_{\phi\to K^+ K^-}\mathcal{B}_{\eta\to\gamma\gamma}$ is the product of the branching fractions cited from the world average values~\cite{pdg}, and $\epsilon$ is the detection efficiency.

In order to obtain the best possible estimate for the detection efficiencies, the signal MC samples are corrected in two aspects:
\begin{enumerate}[(i)]
  \item The track helix parameters~\cite{guoyuping_helix} are corrected to reduce the difference of the kinematic fit $\chi^2_{\rm 4C}$ between the data and MC sample, where the correction factors are obtained from a clean control sample of $\psi(3686)\to K^+ K^- \pi^+ \pi^-$ decay.
  \item Taking into account E1 transition effects on the lineshapes of $\chi_{cJ}$ mesons generated with the Breit-Wigner functions, a weighting factor, $(\frac{E_{\gamma1}}{E_{\gamma10}})^3$~\cite{fanjz_egam3} is applied to the $M_{\phi K^+ K^- \gamma \gamma}$ spectra, where $E_{\gamma1}$ is the radiative photon's energy in the rest frame of the $\psi(3686)$ meson without detector reconstruction effects, and $E_{\gamma10}$ is the most probable transition energy,
\begin{equation}
  \label{Amp}
   E_{\gamma10} = \frac{E^2_{\rm cms} -  m^2_{\chi_{cJ}}}{2E_{\rm cms}}.
\end{equation}
Here $m_{\chi_{cJ}}$ are the nominal masses of the $\chi_{cJ}$ mesons~\cite{pdg} and $E_{\rm cms}$ is the center-of-mass energy of 3.686~GeV.
\end{enumerate}
The detection efficiencies are determined to be $(5.30\pm0.02)\%$, $(6.77\pm0.03)\%$, and $(6.62\pm0.03)\%$ for $\chi_{c0}$, $\chi_{c1}$, and $\chi_{c2} \to \phi \phi \eta$ decays, respectively.

\section{\boldmath SYSTEMATIC UNCERTAINTIES}\label{Sys_err}

The sources of systematic uncertainty include the total number of $\psi(3686)$ events, the MDC tracking efficiency, PID efficiency, photon detection efficiency, $\eta$ and $\phi$ mass requirements, kinematic fit, fit procedure, peaking background estimation, and cited branching fractions.

The total number of $\psi(3686)$ events is $N_{\psi(3686)} = (448.1\pm2.9)\times10^{6}$~\cite{psipnumberT}, which is determined by counting hadronic events. The systematic uncertainty is 0.6\%.

The control samples of $J/\psi \to K_S^0 K^{\pm} \pi^{\mp}, ~K^0_S \to \pi^+ \pi^-$ decays~\cite{kaon_track_sys} has been used to investigate the MDC tracking efficiency, and the difference of 1\% per $K^{\pm}$ track between the data and MC simulation is assigned as the systematic uncertainty. By means of the same control sample, the uncertainty due to PID efficiency is estimated to be also 1\% per $K^{\pm}$ track. The systematic uncertainty from the photon detection efficiency is determined to be 1\% per photon utilizing a control sample of $J/\psi \to \rho^0 \pi^0$ with $\rho^0 \to \pi^+ \pi^-$ and $\pi^0 \to \gamma \gamma$~\cite{photon_sys}.

The systematic uncertainty arising from the $\eta$ ($\phi$) mass requirement is evaluated by changing the mass resolution and shifting the mass window.
In the nominal fit, the $\eta$ signal shape is described as the signal MC simulation convolved with a Gaussian function, and the $\phi$ signal shape is modeled as a $P$-wave Breit-Wigner function convolved with a Gaussian function. Alternative fits are performed by modeling the $\eta$ signal shape with the signal MC simulation, changing the width of the Gaussian function for the $\phi$ signal shape to that obtained from the signal MC sample, and varying the $\eta$ ($\phi$) mass window by the respective mass resolution obtained in the fit to the signal shape. The difference of the efficiency of the $\eta$ ($\phi$) mass requirement between the data and MC sample is taken as the systematic uncertainty from the $\eta$ ($\phi$) mass requirement.

In the nominal analysis, the track helix parameters for charged tracks from signal MC samples are corrected to improve the agreement between the data and MC simulation. The alternative detection efficiency is obtained with no correction on the track helix parameters, and the difference is assigned as the systematic uncertainty associated with the kinematic fit.

The sources of systematic uncertainty from the fit procedure include the signal shape, background shape, and the fit range.

\begin{enumerate}[(i)]
  \item In the nominal fit, the $\chi_{cJ}$ signal shape is modeled with the MC simulation. An alternative fit is performed with the MC simulation convolved with a Gaussian function, and the difference of the signal yield is taken as the systematic uncertainty from the $\chi_{cJ}$ signal shape.
  \item Different order Chebychev functions instead of a constant are used in the alternative fits to describe the background. The largest difference of the signal yield is assigned as the systematic uncertainty from the background shape.
  \item The fit ranges are varied from [3.3, 3.6]~GeV/$c^{2}$ to [3.25, 3.61] or [3.35, 3.6]~GeV/$c^{2}$. The largest difference of the signal yield is taken as the systematic uncertainty associated with the fit range.
\end{enumerate}
 The quadratic sum of the above three systematic uncertainties is taken as the systematic uncertainty from the fit procedure.

In the nominal fit, events in the $\eta$ sideband and 2-D $\phi$ sideband regions are used to estimate contributions from peaking background sources with no $\eta$ signal and only one $\phi$ signal, respectively. Alternative fits are performed by varying the $\eta$ and $\phi$ sideband regions by the mass resolution, and the largest difference of the signal yield is taken as the corresponding systematic uncertainty. The quadratic sum of the two cases is taken as the systematic uncertainty from peaking backgrounds.

 The uncertainties associated with the branching fractions of $\psi(3686)\to\gamma \chi_{cJ}$, $\phi \to K^+ K^-$ and $\eta \to \gamma \gamma$ are estimated from the world average values~\cite{pdg}. The systematic uncertainty due to the trigger efficiency is negligible according to the studies in Ref.~\cite{trigger}.

The total systematic uncertainty on the measured branching fractions for $\chi_{cJ}\to\phi \phi \eta$ decays
 is the quadratic sum of each individual contribution, as summarized in Table~\ref{sysfitetaptotal}.

\begin{table}[htp]
\begin{center}
\caption{Relative systematic uncertainties on the measured branching fractions of $\chi_{cJ}\to \phi \phi \eta$ decays~(in percent).}
\label{sysfitetaptotal}
\setlength{\extrarowheight}{0.9ex}
\renewcommand{\arraystretch}{0.9}
\vspace{0.2cm}
\begin {tabular}{p{4.5cm}m{1.0cm}<{\centering}m{1.0cm}<{\centering}m{1.0cm}<{\centering}}
\hline\hline
Source & $\chi_{c0}$ & $\chi_{c1}$ & $\chi_{c2}$ \\
\hline
$N_{\psi(3686)}$           & 0.6 & 0.6 & 0.6  \\
MDC tracking               & 4.0 & 4.0 & 4.0  \\
PID                        & 4.0 & 4.0 & 4.0  \\
Photon detection           & 3.0 & 3.0 & 3.0  \\
$\eta$ mass requirement    & 0.2 & 0.2 & 0.2  \\
$\phi$ mass requirement    & 0.2 & 0.2 & 0.2  \\
Kinematic fit              & 1.3 & 1.4 & 0.7  \\
Fit procedure              & 1.0 & 0.9 & 1.2  \\
Peaking backgrounds        & 1.3 & 0.9 & 0.9\\
Cited branching fractions  & 2.9 & 3.1 & 2.9 \\
\hline
Total                      & 7.4 & 7.4 & 7.3 \\
\hline\hline
\end{tabular}
\vspace{-0.2cm}
\end{center}
\end{table}

\section{\boldmath{RESULTS AND DISCUSSION}}
The measured branching fractions of $\chi_{cJ} \to \phi \phi \eta$ decays are summarized in Table~\ref{result_branching_chicj}, where the first uncertainties are statistical, and the second are systematic.

\begin{figure}[tb]
\centering
   \includegraphics[width=0.45\textwidth]{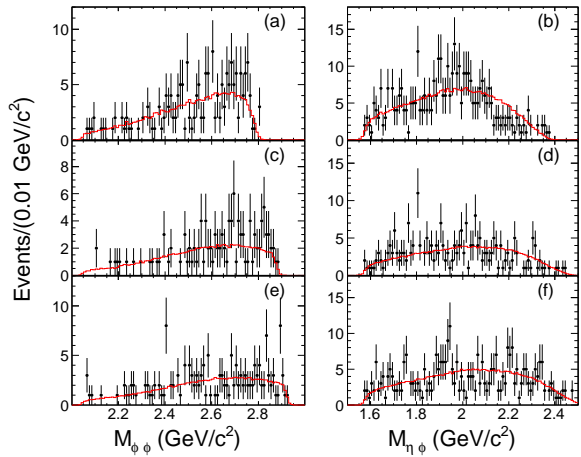}
  \caption{Comparison of the $M_{\phi \phi}$ and $M_{\eta \phi}$ spectra for data (dots with error bars) and signal MC (solid red line) samples within (a)(b) $\chi_{c0}$, (c)(d) $\chi_{c1}$, and (e)(f) $\chi_{c2}$ signal regions.}
\label{com_mphiphi_data}
\end{figure}

Figure~\ref{com_mphiphi_data} shows the projections on the $M_{\phi\phi}$ and $M_{\eta\phi}$ spectra. There are two combinations of $M_{\eta\phi}$ for each event. Compared with those from the signal MC samples, some excesses in data are observed. However, considering the limited statistics, it is hard to draw a conclusion that intermediate states appear in $\chi_{cJ}\to \phi \phi \eta$ decays. Perhaps in the future, utilizing more data samples, it would be worthwhile to combine other $\phi$ and $\eta$ decay modes, such as $\phi\to \pi^+\pi^- \pi^0$, $\phi\to K_S^0 K_L^0$, and $\eta \to \pi^+ \pi^- \pi^0$ decays, to perform a partial wave analysis of $\chi_{cJ}\to \phi\phi\eta$ decays, so that we can make clear conclusions on the existences of intermediate states.

\begin{table}[tb]
\begin{center}
 \caption{Summary of the resulting branching fractions for $\chi_{cJ}\to \phi \phi \eta$ decays.}
  \label{result_branching_chicj}
  \setlength{\extrarowheight}{1.0ex}
  \renewcommand{\arraystretch}{1.0}
  \vspace{0.2cm}
  \begin{tabular}{p{2.0cm}m{3.5cm}<{\centering}}
  \hline\hline
Mode & $\mathcal{B}(\times 10^{-4})$   \\\hline
$\chi_{c0} \to \phi \phi \eta$ & $8.41\pm0.74\pm0.62$   \\
$\chi_{c1} \to \phi \phi \eta$ & $2.96\pm0.43\pm0.22$   \\
$\chi_{c2} \to \phi \phi \eta$ & $5.33\pm0.52\pm0.39$  \\
  \hline\hline
  \end{tabular}
  \vspace{-0.2cm}
  \end{center}
\end{table}

\section{\boldmath SUMMARY}\label{summary}

In brief, the decays $\chi_{cJ} \to \phi \phi \eta$ have been measured for the first time through $\psi(3686)$ radiative decays, based on $4.48\times 10^8$ $\psi(3686)$ events collected with the BESIII
detector. The resulting branching fractions are $(8.41\pm0.74\pm0.62)\times10^{-4}$, $(2.96\pm0.43\pm0.22)\times10^{-4}$, and
$(5.33\pm0.52\pm0.39)\times10^{-4}$ for $\chi_{c0,1,2}\to\phi\phi \eta$ decays, where the first and second uncertainties are statistical and systematic, respectively. At the present level of statistics, no obvious resonant structure is observed in the $M_{\phi \phi}$ or $M_{\eta \phi}$ spectra.

  \section*{\boldmath ACKNOWLEDGMENTS}

  The BESIII collaboration thanks the staff of BEPCII and the IHEP computing center for their strong support. This work is supported in part by National Key Basic Research Program of China under Contract No. 2015CB856700; National Natural Science Foundation of China (NSFC) under Contracts Nos. 11805037, 11625523, 11635010, 11735014, 11822506, 11835012; the Chinese Academy of Sciences (CAS) Large-Scale Scientific Facility Program; Joint Large-Scale Scientific Facility Funds of the NSFC and CAS under Contracts Nos. U1832121, U1632107, U1532257, U1532258, U1732263, U1832207; CAS Key Research Program of Frontier Sciences under Contracts Nos. QYZDJ-SSW-SLH003, QYZDJ-SSW-SLH040; 100 Talents Program of CAS; INPAC and Shanghai Key Laboratory for Particle Physics and Cosmology; ERC under Contract No. 758462; German Research Foundation DFG under Contracts Nos. Collaborative Research Center CRC 1044, FOR 2359; Istituto Nazionale di Fisica Nucleare, Italy; Koninklijke Nederlandse Akademie van Wetenschappen (KNAW) under Contract No. 530-4CDP03; Ministry of Development of Turkey under Contract No. DPT2006K-120470; National Science and Technology fund; STFC (United Kingdom); The Knut and Alice Wallenberg Foundation (Sweden) under Contract No. 2016.0157; The Royal Society, UK under Contracts Nos. DH140054, DH160214; The Swedish Research Council; U. S. Department of Energy under Contracts Nos. DE-FG02-05ER41374, DE-SC-0010118, DE-SC-0012069; University of Groningen (RuG) and the Helmholtzzentrum fuer Schwerionenforschung GmbH (GSI), Darmstadt.


\end{document}